\documentclass[twocolumn,aps,pra]{revtex4}
\usepackage{amsmath,amssymb,mathrsfs,bm}
\usepackage{graphicx,dcolumn,times}
\newcommand{\enquote}{\textit}

\begin{document}
\title{High efficiency photonic storage of single photons in cold atoms}

\author{Jianfeng Li$^1$\footnotemark[2], Yunfei Wang$^1$\footnotemark[2],
Shanchao Zhang$^2$\footnote[2]{These authors contributed equally to this work.},
Junyu He$^1$, Aiqin Cheng$^1$, Hui Yan$^1$\footnotemark[1], and
Shi-Liang Zhu$^{3,1,4}$\footnote[1]{email: yanhui@scnu.edu.cn; slzhu@nju.edu.cn}}

\affiliation{$^1$Guangdong Provincial Key Laboratory of Quantum Engineering and Quantum Materials, SPTE, South China Normal University, Guangzhou 510006, China\\
$^2$Department of Physics, The Hong Kong University of Science and Technology, Clear Water Bay, Kowloon, Hong Kong,China\\
$^3$National Laboratory of Solid State Microstructures, School of Physics, Nanjing University, Nanjing 210093, China\\
$^4$Synergetic Innovation Center of Quantum Information and
Quantum Physics, University of Science and Technology of China,
Hefei 230026, China}
\date{\today}

\begin{abstract}
An efficient quantum storage is highly desired for quantum information
processing. As indicated by certain applications, a universal 
quantum storage is required to have a storage efficiency above 50\% 
to beat the no-cloning limit. Although significant progress has been achieved 
in improving various quantum storage, the best storage efficiency of 
single photons is still below this criteria. By integrating a highly 
controllable single photon source with an optimized quantum storage, 
here we demonstrate an optical storage of single photons with storage efficiency of 65\% 
in a cold atomic ensemble based on electromagnetically induced transparency. 
Meanwhile, the nonclassical characteristics of our storage are verified through the
well-maintained nonclassical and single photon nature of the retrieved single photons.
\end{abstract}

\maketitle

\section{Introduction}
A quantum storage (QS) that records quantum information,
generally encoded in the quantum state of a qubit, for later
on-demand retrieval is indispensable to the development of quantum
information processing \cite{LQCRev,Kuzmich2005,
QInternet,QRep2011,QRev2013,Duan}.
A single photon, the fundamental energy quanta of light, is the most
promising candidate for the flying qubit that transmits the quantum
information between quantum nodes. Therefore, an available QS of
single photon lies at the heart of building up the large scale
quantum network. Since the first demonstration of storing the
quantum state of light in the material \cite{classicallimit}, many
QS schemes based on coherent light-matter interaction have been
proposed and successfully demonstrated
\cite{QMemRev2009,QMemRev2016}, such as electromagnetically
induced transparency (EIT)
\cite{Kuzmich2005,classicallimit,EIT2007Novikova,EIT2008Squeez,EITBEC,
EIT2011Pan,EIT2012Du,EIT2013Yu,EIT2013Ding,EIT2016Yu,EIT2016Halfm,
EIT2016Wang,EITRMP},
photon echo
\cite{SE69GEM2010,AFC2014Gisin,AFC2015Tittel,SE87GEM2016} and
off-resonance Raman interaction
\cite{Ram2010Walmsley,Ram2011Rempe,Ram2015Guo,Ram2016Nunn}.

An ideal QS must be capable of releasing a qubit in a lossless and
quantum way after storing it for a controllable time. 
While certain loss and noises are inevitable in reality, 
a univeral QS of photonic qubit must surpass the no-cloning limit 
\cite{SE87GEM2016} with storage efficiency (SE) above 
the threshold of 50\% in many quantum applications. 
More practically, to implement the long-distance quantum communication,
a larger SE would significantly improve the transmission rates
\cite{QRep2011,QRev2013,SE69GEM2010}.

In the past a few years, extraordinary progress have been made in
improving the performance of QS of nonclassical light
\cite{EIT2008Squeez,SE69GEM2010,EIT2012Du,EIT2011Pan,EITBEC,
EIT2013Ding,AFC2014Gisin,AFC2015Tittel,Ram2015Guo} and coherent
light \cite{EIT2007Novikova,EIT2013Yu,EIT2016Halfm,EIT2016Yu,EIT2016Wang,
SE87GEM2016,Ram2016Nunn,Ram2010Walmsley,Ram2011Rempe}. 
Up to date, although the highest SE of above 90\% has been successfully
demonstrated with laser pulse \cite{EIT2016Yu} and the high SE 
of weak coherent pulse at single-photon level reaches 73\%\cite{SE87GEM2016}
, the best SE of true single photons \cite{EITBEC,EIT2012Du,AFC2014Gisin,
AFC2015Tittel,Ram2015Guo,AFC2015Tele,SE2015Pan,EIT2014OAM} 
is still lower than the threshold. The main obstacle to realizing a higher
SE of single photons lies at the difficulty of both optimizing the
QS and integrating the QS with the versatile single photon
sources. 

In this paper, we report a storage of heralded single photons
with SE of 65\% that is above the threshold for the first time.
SE dependence on the controllable parameters of the storage, 
the optical depth and the control laser intensity, are explored and 
the strategy of optimizing these parameters to obtain the best SE
are presented in detail. To achieve this high storage efficiency, 
we firstly built up a highly controllable single photons source 
with photon wavelength matching the storage. Secondly, to achieve 
the high optical depth, a magneto-optical trap (MOT) with 
large trapping volume is designed for the sake of better
system robustness and greater duty cycle while comparing to 
the setups of spatial \cite{EIT2012Du} and temporal \cite{EIT2016Yu} 
dark MOT. The dephasing rate of our storage is minimized by 
compensating the residual magnetic field. Photon noises are quelled by 
implementing both the spatial and spectrum filters. 
Eventually, we succeed in realizing an efficient storage of 
single photons and our work is an important step towards 
the practical applications of QS.

\begin{figure}[htpb]
\begin{center}
\includegraphics[width=8.5cm]{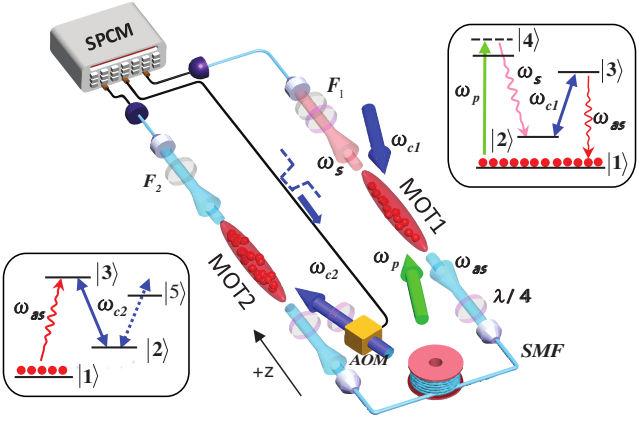}
\caption{\label{fig:1} (color online). Experimental setup:
In MOT1, with the presence of a pump ($\omega_{p}$) and
coupling ($\omega_{c1}$) lasers, counter-propagating Stokes
($\omega_{s}$) and anti-Stokes ($\omega_{as}$) paired photons are
produced via the spontaneous four-wave-mixing (sFWM) process.
The anti-Stokes photons ($\omega_{as}$) are sent to MOT2
via a 100m long polarization-maintained single mode fiber (SMF).
By switching off and on the control light ($\omega_{c2}$) with
an acousto-optic modulator (AOM), the photon can be stored
into and readout from the electromagnetically
induced transparency (EIT) based optical storage (OS) that
operates in MOT2. The detailed $^{85}Rb$ energy levels shown
in the right inset for sFWM configuration and
in the left inset for the EIT scheme are chosen as
$|1\rangle=|5S_{1/2},F=2\rangle$,
$|2\rangle=|5S_{1/2},F=3\rangle$,
$|3\rangle=|5P_{1/2},F=3\rangle$,
$|4\rangle=|5P_{3/2},F=3\rangle$, and
$|5\rangle=|5P_{1/2},F=2\rangle$.
$\lambda/4$: Quarter wave plates;
SPCM: Single-photon counter module;
$F1$ and $F2$: Multi-level Fabry-Perot filters.}
\end{center}
\end{figure}
\section{Experimental setup and single photon generation.}
The experimental setup is sketched in Fig.~\ref{fig:1}.  It
consists of a source that generates heralded single photons from a
cigar-shaped MOT (MOT1) and a OS that operated in a second
cigar-shaped MOT (MOT2). To integrate the single photon source
with our EIT-based OS in MOT2 of $^{85}$Rb atoms, we use the same
atoms with proper energy level configuration in MOT1 to generate
heralded single photons via the backward spontaneous four-wave
mixing (sFWM) process, which automatically matches the single
photon wavelength to the working wavelength of OS. l
The whole experiment runs periodically with a repetition rate of
100~Hz. In each cycle, after 9.5~ms loading time, atoms in both
MOT1 and MOT2 are prepared in the same ground state $\left|
1\right\rangle$. Following that is a 0.5~ms single-photon
generation and storage window.

Significantly different from the storage process in the demonstrations
using laser pulse\cite{EIT2007Novikova,EIT2016Yu, EIT2016Wang,
SE87GEM2016,Ram2016Nunn,Ram2010Walmsley,Ram2011Rempe},
storage of the heralded single photon in MOT2 is triggered by the
detection of spontaneously generated Stokes photon, which
requires the atomic ensemble property to stable during the storage
window and hence our OS has an optimized duty cycle of 5\% with
the storage window of 500$\mu s$. During this window, once a Stokes
photon is detected, its paired anti-Stokes photon is projected
into a single photon state as a heralded single photon and ready
to be stored in MOT2.

\subsection{Heralded single photon generation.}
To generate the heralded single photons, backward spontaneous
four-wave-mixing (sFWM) configuration is implemented in MOT1 with
longitudinal length $L_1=1.5~cm$, which includes a
counter-propagating pump/coupling beam pair. The pump laser
($780~nm,\sigma^{-},\omega_p$) is a focusing beam with its
focus point locating the center of MOT1 and has the
frequency blue detuned from the transition
$|1\rangle\leftrightarrow|4\rangle$ by 80~MHz. The coupling laser
($795~nm,\sigma^{+},\omega_{c1}$) is a nearly collimated Gaussian
beam with a $1.8~mm$ $1/e^2$ diameter and has  a frequency
resonant with the transition $|2\rangle\leftrightarrow|3\rangle$.
Phase-matched Stokes ($\omega_s$) and anti-Stokes ($\omega_{as}$)
paired photons are produced along the longitudinal axis and
coupled into two opposing single-mode fibers (SMFs). The two SMFs
have mirrored spatial modes that both focus at the MOT1 center
with a  0.24~mm $1/e^2$ diameter, which matches the spatial mode
of the probe laser beam for measuring the longitudinal optical
depth of MOT1. The angle between the Stokes/anti-Stokes axis and
the pump/coupling axis is chosen to be $2.5^{o}$, which provides
spatial separation of the weak Stokes and anti-Stokes fields from
the strong pump and coupling fields. After a narrowband
Fabry-Perot (FP) filter ($F_{1}$) with isolation ratio of 38~dB,
the Stokes photons are detected by a single-photon counter module
(SPCM, Perkin Elmer, SPCM-AQ4C) with a dark count rate of 500
counts per second.

\begin{figure}
\begin{center}
\includegraphics[width=8.5cm]{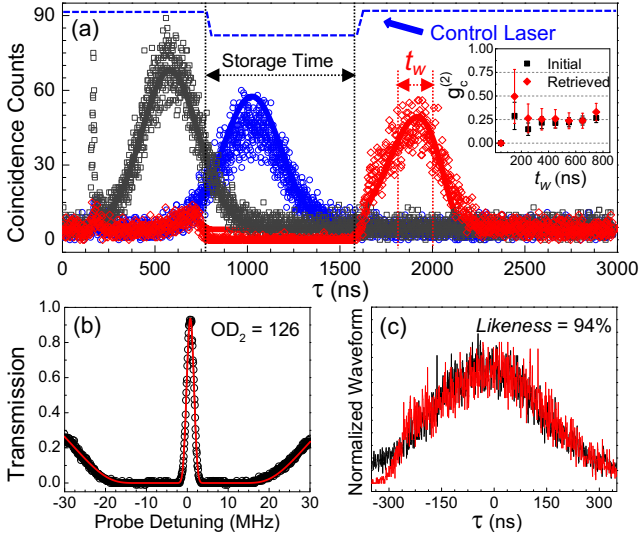}
\caption{\label{fig:2} (color online).
(a) Typical temporal waveforms with total photon generation time of 3000~s.
The measured initial, EIT-slowed and retrieved waveforms of heralded
single photons are shown as black open square, blue open circle and
red open diamond, while the theoretical curves are depicted as
black, blue, and red lines, respectively. The dashed blue
line indicates the temporal waveform of the control laser.
The inset figure shows the typically measured conditional 
auto-correlation function $g_{c}^{(2)}$ of the initial (red diamond)
and retrieved photons (black square) when the measurement window 
$t_W$ is changed around the peak of the respective Gaussian shape
 waveform. The error bars are obtained by considering the 
Poisson statistics of the photon counting.
(b) The typical EIT transmission profile of the MOT2 with
optical depth $OD_2=126$. Black open circles are
experimental data and the red line is the best fitted theoretical curve.
(c) Waveform likeness between the normalized temporal waveform of
heralded single photon before and after storage.}
\end{center}
\end{figure}

A typical temporal waveform of the heralded single photons
that are used in all the following OS experiments is shown in
Fig.~\ref{fig:2}(a) (black open square),  where the
main wavepacket has a full width at half maximum (FWHM)
around 400~ns. The sharp spike in the left of Fig. 1(a) is the
optical precursor \cite{SPhPrecursor} that always propagates with
light speed $c_0$. Taking into account the total channel efficiency
(2.7\%) and experiment duty cycle (5\%), the single photon
generation rate in the source is about 6400 photons per second. 
Here, the channel efficiency includes all the fiber coupling 
efficiency, the efficiency of optical filters and 
the detector efficiency of SPCM.

To produce single photons with above waveform, we set the
optical depth of MOT1 $OD_1=100$ and Rabi frequency of the
coupling laser $\Omega_{c1}=3.5 \gamma_{13}$, with
$\gamma_{13}=2\pi\times$3.0~MHz being the electric dipole
relaxation rate between states $|1\rangle\leftrightarrow|3\rangle$.
The focused pump laser beam ($\Omega_{p}$) is achived by
a plano-convex lens instead of the complicated
spatial light modulator setup\cite{sFWMSLM2015Du}.
Its Rabi frequency is designed to have a spatial distribution of
$\Omega_{p}(z)=\Omega_{p0}f_{p}(z)$,
where $\Omega_{p0}=0.5\gamma_{13}$,
$f_{p}(z)=e^{-z^{2}/(4.16 mm)^{2}}$ and $z=0$ indicate 
the center of MOT1 (See supplement materials). This spatial 
profile of pump laser makes the main part of the heralded single
photon waveform a Gaussian shape as designed. The theoretical
curve, shown in Fig.~\ref{fig:2}(a) (black line), can be
calculated according to the following
equation\cite{sFWMSLM2015Du}:
\begin{equation}
\begin{split}\label{Eq:sFWM}
\psi_{as}(\tau)=\kappa_{0}V_{g1}f_{p}(L_1/2-V_{g1}\tau)e^{-i\omega_{as}\tau},
\end{split}
\end{equation}
where $\psi_{as}(\tau)$ is the wavepacket of heralded single
photons with $\tau$ as the relative time defined respect to the
time instant when a Stokes photon is detected in each experiment
period, $\kappa_{0}$ is the nonlinear coupling strength of sFWM
process, and $V_{g1}=\Omega_{c1}^{2} L_1/(2 OD_1 \gamma_{13})$ is
the group velocity of the heralded single photon in MOT1.

The quantum nature of the generated photon can be verified with a
Hanbury, Brown and Twiss interferometer that shows the
non-separability of a single photon. A pure heralded single photon 
state promises $g_{c}^{(2)}$=0, while a two-photon state 
gives $g_{c}^{(2)}$= 0.5. In our experiment, a finite measurement 
window $t_W$ is taken to get a resonable evaluation of $g_{c}^{(2)}$, 
as shown in the inset figure in \ref{fig:2}(a). The definition of 
$t_W$ of generated photons is respect to the Gasussian peak position 
of its own waveform. For reference, only the $t_W$ of retrieved photon 
is depicted in \ref{fig:2}(a). With a reasonable windown,
the generated shows a stable value $g_{c}^{(2)}<0.5$. 
At a large enough window of $t_W=800 ns$, we have 
$g_{c}^{(2)}=0.26\pm0.05$ with uncertainties derived from the
Poisson statistics of the recorded photon counts, which strongly 
suggests a near-single-photon characteristic of generated photons.
The nonclassical correlation betwen Stokes photon and 
anti-Stokes photon can be measured by the quantity of 
$R_{CS}=g^2_{s,as}/(g_{s,s}g_{as,as})$\cite{QRep2011}, 
which means how many times the Cauchy-Schwarz inequality is violated. 
For our photon source, we have obtained a $R_{CS}\approx54$ 
around the peak of the main waveform.

\subsection{Optical storage setup.}
A pair of opposing SMFs is setup in a way similar to that used
for Stokes/anti-Stokes collection in MOT1 to deliver and collect
the single photons through MOT2
with a longitudinal length of $L_2=2.8~cm$. In each cycle, the
quadruple gradient magnetic field in MOT2 is shut down in the last
1~ms of the atoms loading and state preparation period. During the
storage window, a strong control laser ($\omega_{c2}$) with a
4.7~mm $1/e^2$ diameter  and the same frequency as $\omega_{c1}$
is switched on. The control laser ($\Omega_{c2}$) can be switched
off and on quickly with a switching time of 70~ns by an
acousto-optic modulator (AOM) driven by a digital waveform
generator (Rigol DSG815), which controls the experiment timing of
storing and and read-out the single photon. The angle spaned by
the single photon axis and the control laser axis
is also chosen to be $2.5^{o}$.
After passing through MOT2 and another 37~dB
narrowband FP filter ($F_{2}$), the heralded single photons are
detected by a second SPCM (Perkin Elmer SPCM-AQRH-16FC) with a
dark count rate of 25 counts per second. At last, the temporal
waveforms of heralded single photons are recorded by a
time-to-digital converter (Fast Comtec P7888) with a time bin
resolution of 1~ns.

\section{Storage efficiency optimization and noise suppression.}
SE of the storage of a single photon is defined as the
ratio between the probability of successfully retrieving the
single photon and the probability of a single photon being
sent into the OS, i.e.,
\begin{equation}
SE=\frac{\int |\psi_{out}(\tau)|^{2} d\tau}{\int
|\psi_{in}(\tau)|^{2}d\tau},
\end{equation}
where $\psi_{in}$ ($\psi_{out}$) is the wavepacket of the single
photons before (after) the storage. The theoretical probabilities
$|\psi_{out}(\tau)|^{2}$ and $|\psi_{in}(\tau)|^{2}$ can be
derived by numerically solving Eqs.(3) detailed below. More
practically, the well separated precursor part in $\psi_{in}$ is
excluded from the SE calculation because it cannot be slowed down
and stored in the EIT scheme, which typically contributes less
than 2\% to the resulted SE.

\subsection{Theoretical model of the EIT based storage}
For guidance of the experimental optimization of the our OS,
theoretical models\cite{DSP2001Lukin,QMemOptTheo2007}
based on the Maxwell-Schrodinger equation of single photon field
and the optical  Bloch equation of atomic ensembles
are implemented to describe our storage,
which is detailed below and has been demonstrated
to work well with the EIT-based storage\cite{EIT2016Yu}:
\begin{eqnarray}
\begin{array}{ll}
&(\partial_{\tau}{} + c_0\partial_{\tilde{z}}{})\tilde{\varepsilon}_{as}= ig\sqrt{N}\tilde{P} \\
&\partial_{\tau}{\tilde{P}}= -\gamma_{13}\tilde{P}+ \frac{ig\sqrt{N}}{2}\tilde{\varepsilon}_{as}+\frac{i}{2}\tilde{\Omega}_{c2}\tilde{S} \\
&\partial_{\tau}{\tilde{S}}= -\gamma_{12,eff}\tilde{S}+\frac{i}{2}\tilde{\Omega}^*_{c2}\tilde{P}
\end{array}
\end{eqnarray}
where $\tilde{\varepsilon}_{as}$ is the time- and position-
dependent slow  varying envelop of the heralded single photon's
quantum field and $\tilde{\Omega}_{c2}$ is the Rabi frequency of
the control laser. $\tilde{P}$ and $\tilde{S}$ are the slow
varying envelope of the collective polarization of
$|1\rangle\leftrightarrow|3\rangle$ coherence and
$|1\rangle\leftrightarrow|2\rangle$ coherence, respectively. $g$
is the photon-atom coupling strength and can be obtained from the
relation $OD_2=g^2NL_2/(\gamma_{13}c_0)$. $N$ is the total atom
number in the interaction volume with the assumption that atoms
are distributed uniformly over MOT2.
$\gamma_{12,eff}=\gamma_{12}+\gamma_{13}(\beta\Omega_{c2})^{2}/(4\Delta_{s}^{2})$
is the effective dephasing rate of ground states
$|1\rangle\leftrightarrow|2\rangle$, where $\beta=\sqrt{37/50}$ is
the Clebsch-Gordon coefficient ratio of
$|2\rangle\rightarrow|5\rangle$ transition to the control laser
transition, and $\Delta_{s}=2\pi\times361.6$~MHz is the hyperfine
excited state splitting between states $|5\rangle$ and $|3\rangle$
of $^{85}Rb$ D1 lines. Here, the extra dephasing rate is
introduced by the intense control laser that off-resonantly
couples ground state $|2\rangle$ to the excited state $|5\rangle$,
which has spontaneous decay rate of $2\gamma_{13}$. With the
anti-Stokes entrance position as $\tilde{z}=0$ in MOT2, we thus have
$\psi_{in}(\tau)=\psi_{as}(\tau)$ and
$\tilde{\varepsilon}_{as}(\tau,\tilde{z}=0)=\psi_{as}(\tau)e^{i\omega_{as}\tau}$.
After numerically solving Eq. (3) we obtain
$\tilde{\varepsilon}_{as}(\tau,L_2)$ and then $\psi_{out}(\tau)$
in the definition of SE in Eq. (2) can be derived from the
relation
$\psi_{out}(\tau)=\tilde{\varepsilon}_{as}(\tau,L_2)e^{-i\omega_{as}\tau}$.

\subsection{Optimization of the single photon storage}
Based on the above theorectical mode, the SE of our OS
is mainly determined by the optical depth ($OD_2$) 
of OS, the Rabi frequency ($\Omega_{c2}$) of control laser
and the dephasing rate ($\gamma_{12}$) of the coherence between
the two states $|1\rangle$ and $|2\rangle$.
We thus adopt the following protocol to optimize our OS:
after preparing MOT2 with a certain optical depth $OD_2$,
we reduce the value of $\gamma_{12}$ in a way that will be
detailed below. $\Omega_{c2}$ is then scanned to optimize the SE
of storing the single photon with a specific waveform.
By repeating this procedure for another value of $OD_2$,
we can obtain the optimal SE for all the experimentally
reachable $OD_2$, $\gamma_{12}$ and $\Omega_{c2}$.

To achieve high enough optical depth $OD_2$, we increase
the beam size of the trapping and repumping laser in MOT2 and
hence expand the MOT2 trapping volume by more than 3 times.
Eventually,  we can tune the $OD_2$ between 30 to 180 by adjusting
the intensity of the repumping laser. The value of $OD_2$ is
measured separately by fitting the EIT transmission curve of a
weak probe laser beam passing through MOT2 with the same spatial
and polarization mode as that of the heralded single photon. The
probe laser frequency is swept around the resonance of atomic
transition $|1\rangle\leftrightarrow|3\rangle$ with the control
laser being continuously present. A typical EIT transmission curve
with $OD_2=126$ is shown in Fig.\ref{fig:2}(b).

With all the other parameters fixed, the SE is higher for a smaller
$\gamma_{12}$. In our setup, finite inhomogeneous stray
magnetic fields and atoms' residual thermal motion are two
main contributions to $\gamma_{12}$. In order to minimize
the magnetic field effect, we switch off the quadruple magnetic field
and turn on three pairs of compensation Helmholtz coils during
the storage window. The atomic motion effect is reduced by
aligning the control laser beam properly with an
optimal angle respect to the single photon path.
The angle we choose at last is $2.5^{o}$, which
is an empirical value in order to restrict the photon noise rate
in the single photon channel to a safe level (90 photons per second)
while maintaining feasible EIT transmission of the single photon.
In addition, this angle is also large enough to effectively suppress
the photon noise generation via 
nonlinear optical processes\cite{EIT2016Yu}.

Finally, we measured the SE by scanning $\Omega_{c2}$
with all the other settings unchanged. At each $OD_2$, a proper
value of $\Omega_{c2}$ that result in the optimal SE is obtained.
Notably, to achieve optimal SE, it is desirable for the temporal
waveform of a single photon to possess the time-reversal 
symmetry with the retrieved waveform
\cite{QMemOptTheo2007,EIT2007Novikova,QMemOpt2008}. 
In our OS, this statement is verified by the typically 
greater than 90\% waveform likeness
as shown in Fig.\ref{fig:2}(c), which will be further addressed
below. Moreover, here we demonstrate an efficient way of matching
the temporal mode of the OS to the input single photon waveform by
simply adjusting the intensity of control laser.

\section{Characteristics of the storage.}
A typical optimal storage process and the resulted waveforms
of the heralded single photon are illustrated in Fig.\ref{fig:2}(a)
with $OD_2=126$. The retrieved waveform shown in
Fig.\ref{fig:2}(a) (red open diamonds) has an SE of 62\% with a
storage time of 900~ns. As a reference, the waveform of the
EIT-slowed single photon is also presented in Fig.\ref{fig:2}(a)
(blue open circles) with an efficiency of 78\%. The best fitted
theoretical waveforms in Fig.\ref{fig:2}(a) (blue and red lines) are
achieved with $\Omega_{c2}=7.6\gamma_{13}$ and
$\gamma_{12}=0.004\gamma_{13}$, where $\Omega_{c2}$ and
$\gamma_{12}$ are free parameters during the fitting.
In our theoretical calculation, effective dephasing rate
$\gamma_{12,eff}$ is used, which contains both the intrinsic
$\gamma_{12}$ and the extra dephasing rate induced by
the off-resonant coupling between states
$|2\rangle\leftrightarrow|5\rangle$.
With the optimized SE, quantum properties of our OS are
characterized by testing the quantum nature of the retrieved
photons. Here, both the single photon nature and waveform
likeness of the retrieved photons are measured.

\subsection{The nonclassical nature of OS.}
The quantum nature of a single photon that can be evaluated by
the auto-correlation function $g_{c}^{(2)}$ is sensitive to the photon
noises introduced by the OS during the storage. In our setup, most of
the photon noises are from the scattering of the intense control
laser beam by various optical elements. To suppress these noises,
besides setting up a finite angle between the control laser beam
and the single photon path, a single mode fiber (SMF) that acts as a
spatial filter is used to collect the retrieved single photon.
Furthermore, a set of temperature stabilized FP etalons are
implemented as spectrum filters. Eventually, at $t_w= 800ns$, 
a safe vale of $g_{c}^{(2)}=0.32\pm0.09$ is measured for 
the retrieved photon, as presented in the figure and inset of 
Fig.\ref{fig:2}(a) (red open diamons and red diamonds). 
Therefore, the single-photon nature of photons during the storage is
well-preserved. On the other hand, the nonclassical correlation 
between retrieved photons and the Stokes photon are also 
preserved by meausring the $R_{CS}\approx29$ at the peak of the 
retrieved photon waveform. With these measurements, 
the nonclassical performance of our OS are well verified.

Waveform likeness is defined as the normalized cross correlation
of two waveforms (See supplements). In the quantum applications 
that rely on the two-photon interference, the photon waveform also 
takes important role\cite{QRep2011, QMemOpt2008}. As shown 
in Fig.\ref{fig:2}(c), the waveform likeness is around 94\%, 
which demonstrates that the waveform is also preserved 
during the storage. We have to point out that, the waveform here 
does not bear any nonclassical nature. Moreover, since a Gaussian 
shape waveform has the intrinsic left-right symmetry, 
this high waveform likness also implys the desired time-reversal 
symmetry \cite{QMemOptTheo2007,EIT2007Novikova,QMemOpt2008}
for an optimal SE.

\begin{figure}
\begin{center}
\includegraphics[width=8cm]{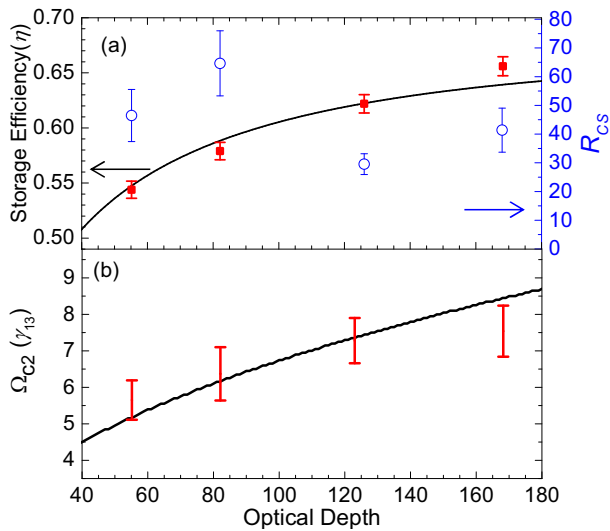}
\caption{\label{fig:3} (Color online).
(a) Storage efficiency optimized for different optical depths ($OD_2$).
The efficiency data refers to the left axis. Error bars are based on 
the Poisson detection statistics and the black line is the theoretical curve.
The nonclassical correlation between retrieved photon
and the Stokes photon are denoted as blue open circle, which 
refers to the right axis.
(b) Experimentally optimized Rabi frequency of control laser
$\Omega_{c2}$ at different $OD_2$. The uncertainty
means a range where the storage efficiencies are indiscernible
within the experimental fluctuations.}
\end{center}
\end{figure}

\subsection{SE dependence on $OD_2$ and $\Omega_{c2}$}
Following the theoretical prediction, the dependence of the 
optimal SE on $OD_2$ are presented in Fig.\ref{fig:3}. 
At all $OD_2$'s, the optimal SE can be achieved with
proper $\Omega_{c2}$ values with the nonclassical properties 
of the retrieved photon preserved well. The red squares and 
blue open circles in Fig.\ref{fig:3}(a) are experiment data 
with error bars propagated from Poisson statistics
of the measured photon counts under each $OD_2$. 
As predicted, SE continues to rise until it reaches 
the highest value around $65\%$ at  $OD_2=168$. 
Theoretical SEs are presented in Fig.\ref{fig:3}~(a) 
as a black line, which agrees well with the experimental data. 
Experimental values of optimal control laser
Rabi frequency $\Omega_{c2}$ at different $OD_2$ are also tested
and depicted in Fig.\ref{fig:3}(b), which show good agreement with
the theoretical curve. Here the uncertainty reflects the range of
$\Omega_{c2}$ that gives a similar optimal SE within the
experimental fluctuations.

\begin{figure}
\begin{center}
\includegraphics[width=8cm]{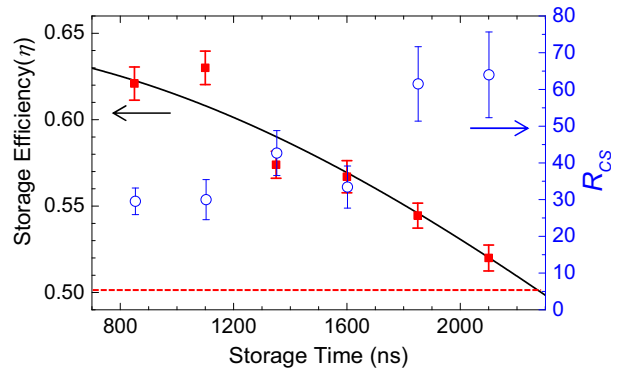}
\caption{\label{fig:4} (Color online). Storage efficiency (SE) with different
storage times. Experimental data are denoted by red squares with
error bars deducted from the Poisson detection statistics.
The black line is the best fitted guideline with the 
Gaussian-decay function and the red dashed line show the SE of 50\%.
The nonclassical property of the retrieved photon 
are denoted as blue open circle that refer to the right axis.}
\end{center}
\end{figure}

\subsection{SE dependence on storage time}
In practical application, a long enough storage time is required. 
Here, the SE dependence on different storage times is
measured and depicted in Fig.\ref{fig:4}. The fractional delay,
which is defined as the ratio of storage time to the initial
waveform FWHM duration, is a useful evaluation in practical
application. It predicts how many quantum unit operations can be
conducted at most during the storage period and has a practically low
limit of 1-bit. In Fig.\ref{fig:4}, the red squares are
experimental SE measured at different storage times with
$OD_2=126$ and $\Omega_{c2}=7.6 \gamma_{13}$, the error bars are
obtained by accounting for the Poisson statistics of measured
photon counts. The black line is the best fit of the data with a
Gaussian decay function\cite{Kuzmich2005,EIT2016Yu}
$e^{-t^{2}/\tau_{0}^{2}}$ that gives our storage coherence time of
$\tau_{0}=4~\mu s$. With a typical SE of $50 \%$, the storage time
is $2.2~\mu s$, which gives a greater than 5-bit fractional delay
with the SE above the 50\%. Although, this is not a large 
value\cite{EIT2013Yu, SE87GEM2016}, a large and promising 
improvement space can be expected with our current setups.

\section{Conclusion.}
In conclusion, we have improved the storage efficiency of heralded single
photons up to $65\%$ in a dense cold atomic ensemble through the
EIT scheme. With an SE of $50\%$, our OS reaches a storage time of
$2.2 \mu s$, which is more than a 5-bit fractional delay. As far
as we know, this is the first demonstration of storing real single
photons with efficiency above the 50\% level. Here, we explore and 
demonstrate an feasible way of integration the single photon source 
and the optical memory, which would pave a way for seeking high 
efficiency storage of real single photon quibit. On the other hand,
our progress would show a add up the confidence to the field 
in developing the practical quantum storage.

\section*{Funding Information}
NKRDP of China (Grants No. 2016YFA0301803 and No. 2016YFA0302800), 
the NSF of  China (Grants No. 11474107, No. 61378012, No. 91636218), 
the GDSBPYSTIT (Grant No.2015TQ01X715), 
the GNSFDYS (Grant No. 2014A030306012), 
the PRNPGZ (Grant No. 2014010).



\newpage
\section{Supplement Materials}
\subsection{Waveform shaping of heralded single photon} 

\begin{figure}[htbp]
\centering
\fbox{\includegraphics[width=8.5 cm]{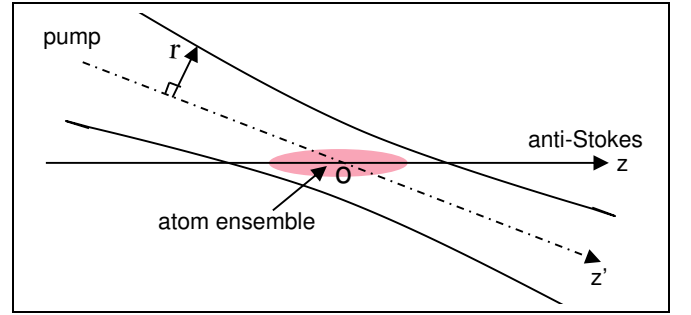}}
\caption{Shematics describing the waveform shaping technique by controlling the pump beam profile. $z$ axis is the longitudinal axis of the atom ensemble (MOT1) and the direction along wich the Stokes and anti-Stokes pair are generated. $z'$ is the propagation direction of pumping light with $r$ as the beam radius of pumping beam in the transverse plane.}
\label{supfig1}
\end{figure}

Here we show the details of how the desired temporal waveform of the heralded single photons can be achieved with the scheme presented in Ref\cite{2}. The temporal waveform of single photon generated by the spontaneous four-wave-mixing (sFWM) process is determined simultaneously by the optical depth($OD_1$), pump laser Rabi frequency $\Omega_{p}$ and coupling laser Rabi frequency $\Omega_{c1}$\cite{1,2}, with all the parameter definition following the main text. According to the discussion in Ref.\cite{1}, our photon source performs in the group delay regime and waveform generated photons are dominated by the group delay effects, which is the starting point to implement this spatial-temporal modulation technique. By intentionally preparing the coupling laser as a near collimated beam with large enough beam size, the generated temporal waveform are mainly determined by the carefully focused pumping beam with a designed spatial distribution of it Rabi frequency along the $z$ axis, as depicted in Fig.\ref{supfig1}. Here, the distribution of pump laser Rabi frequency can be described as $\Omega_p(z)=\Omega_{p0} f_p(z)$, and the envelop of the temporal waveform of emitted single photon can be approximately given as:
\begin{eqnarray}
\begin{array}{ll}
&\tilde{\psi}_{as}(\tau)\approx \kappa_0 \int_{-\frac{L_1}{2}}^{\frac{L_1}{2}}dzf_p(z)
\int{d\omega g_{\chi}^{(3)}(\omega) e^{-i\omega (\tau+\frac{z}{V_{g1}}-\tau_g)}} \\
\end{array}
\end{eqnarray}
with $\kappa_0=\frac{\sqrt{\omega_{s0}\omega_{as0}}}{i4{\pi}c}
\frac{\hbar^2\Omega_{p0}\Omega_{c1}}{\mu_{14}\mu_{23}} 
\chi^{(3)}_0$ as the nonlinear coupling strength of. 
Here, $c$ is the speed of light in vacuum, $\hbar$ is the reduce Planck constant. 
$\omega_{s0/as0}$ is the centeral frequency of Stokes/anti-Stokes photon, 
$\mu_{ij}$ is the dipole matrix element between the state of 
$|i\rangle$ and $|j\rangle$ as the enengy level shown in main text. 
$\chi^{(3)}_0$ is the resonant 3rd order susceptibility and 
$g_{\chi}^{(3)}(\omega)$ is the 3rd order susceptibility spectrum divided by $\chi^{(3)}_0$. 
$V_{g1}=\frac{\Omega_{c1}^{2} L_1}{(2 OD_1 \gamma_{13})}$ is the group velocity 
of the anti-Stokes photons as defined in the main text and $\tau_g=\frac{L_1}{V_{g1}}$. 
$\tau=\tau_{as}-\tau_s$ is the time interval between anti-Stokes photon and Stokes photon. 

With the of integration variable change from $z$ to $t$ as $t=\tau+\frac{z}{V_{g1}}-\tau_g$, 
integration in the space domain in Eq. (S1) changes to integration in the time domain:
\begin{eqnarray}
\begin{array}{ll}
\tilde{\psi}_{as}(\tau)\approx &\kappa_0\int_{\tau-\tau_g}^{\tau}dt f_p(\frac{L_1}{2}+V_{g1}(t-\tau))\\
&\times \int{d\omega g_{\chi}^{(3)}(\omega) e^{-i\omega t}}
\end{array}
\end{eqnarray}
In the group delay regime only the spectrum part near the EIT resonance plays the role in the integration 
about $\omega$ due to the narrow EIT window, 
we thus can safely make the approximation 
$g_{\chi}^{(3)}(\omega) \approx g_{\chi}^{(3)}(0)=1$ and thus Eq. (S2) reduces to
\begin{eqnarray}
\begin{array}{ll}
\tilde{\psi}_{as}(\tau)&\approx \kappa_0\int_{\tau-\tau_g}^{\tau}dtf_p(\frac{L_1}{2}+V_{g1}(t-\tau))\delta(t) \\
&=\kappa_0V_{g1}f_p(\frac{L_1}{2}-V_{g1}\tau)
\end{array}
\end{eqnarray}
Hence, up to here, we briefly show the derivation of the envelope part of Eq.(1) in the main text.
Considering the situation that $f_p(z)$  has a Gaussian shape as below:
\begin{eqnarray}
\begin{array}{ll}
&f_p(z)=e^{-\frac{z^2}{z_0^2}}
\end{array}
\end{eqnarray}
we will directly obtain an Gaussian shape temporal waveform as:
\begin{eqnarray}
\begin{array}{ll}
\tilde{\psi}_{as}(\tau)=\kappa_0V_{g1}e^{-\frac{(\tau-\tau_g/2)^2}{\tau_0^2}}
\end{array}
\end{eqnarray}
with $\tau_0=\frac{z_0}{V_g}=\frac{z_0}{L_1}\tau_g$.

In our setup, a lens is used to focus the pump laser on the center of the atom cloud, 
with the Rayleigh length is much longer than the length of the ensemble, 
the Rabi frequency along the $z$ axis is thus
\begin{eqnarray}
\begin{array}{ll}
&\Omega_p(r,z')=\Omega_p\frac{w_0}{w(z')}e^{-\frac{r^2}{w^2(z')}}\approx{\Omega_pe^{-\frac{r^2}{w_0^2}}}
\end{array}
\end{eqnarray}
Here $w_0$ is the waist of the Gaussian beam, and $z'$ is the direction of propagation.
As shown in Fig.~\ref{supfig1} the angle between the $z$ and $z'$ is chosen to 
be $2.5^\circ$, Thus the we have: 
\begin{eqnarray}
\begin{array}{ll}
&f_p(z)=e^{-\frac{(zsin\theta)^2}{w_0^2}}{\approx}e^{-\frac{z^2}{(w_0/\theta)^2}}
\end{array}
\end{eqnarray}
Then we have
\begin{eqnarray}
\begin{array}{ll}
&\tau_0=\frac{2 OD_1\gamma_{13}}{\Omega_{c1}^2 L_1}z_0
=\frac{2 OD_1 \gamma_{13}}{\Omega_{c1}^2 L_1\theta}w_0
\end{array}
\end{eqnarray}

According to Eq. (S8), we can easily control the width of the gaussian shape biphoton waveform by changing the $OD_1$, $\Omega_{c1}$ or $w_0$. 
In order to a high generation rata, generally we do not change the $OD_1$. 
As we mention in the main text, the parameters of our system are $OD_1=100$,
$\gamma_{13}=2\pi\times3MHz$, $L_1=1.5cm$ and 
$\Omega_{c1}=3.5\gamma_{13}$. 
The focus of the lens we use has a focal length of 8mm and 
the distance between the lens and the center of the ensemble is about 70cm. 
With the waist of the fiber core $w_{fiber}\approx2.5\mu{m}$, 
we have $w_0\approx182\mu{m}$ and thus $\tau_0\approx240ns$.
The full width of half maximum(FWHM) of the shaped biphoton waveform is 
$FWHM=2\tau_0\sqrt{ln2}\approx400ns$, just as we show in the paper. 
The Rabi frequency distribution of the pump laser a long $z$ axis is thus fitted as:
\begin{eqnarray}
\begin{array}{ll}
&f_p(z)\approx{e^{-\frac{z^2}{(w_0/\theta)^2}}}=e^{-\frac{z^2}{(4.16mm)^2}}
\end{array}
\end{eqnarray}
We thus can conclude that a single lens is stable and powerful enough for 
Gaussian type shaping and 
a spatial light modulator(SLM) might be necessary for more complicated shaping. 

\subsection{Measurement of the conditional auto-correlation function} 

\begin{figure}[htbp]
\centering
\fbox{\includegraphics[width=8.5 cm]{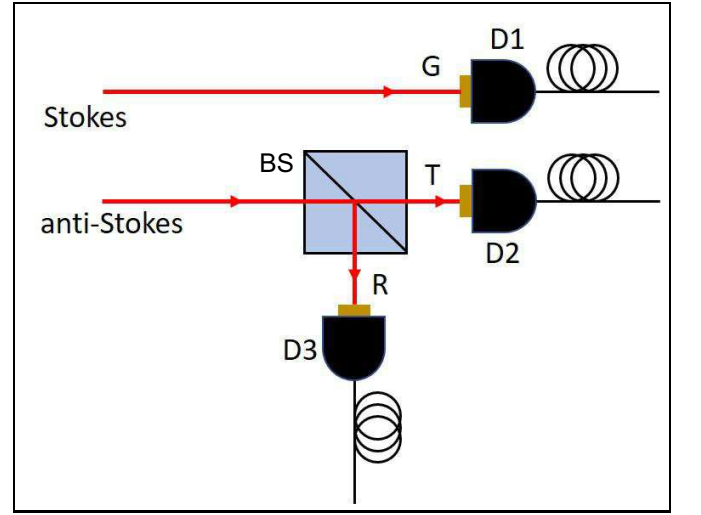}}
\caption{ Experimental setup for measuring the conditional autocorrelation function $g^{(2)}_c$.}
\label{supfig2}
\end{figure}

\noindent
The single-photon quantum nature can be verified by measuring the
conditional auto-correlation function $g_c^{(2)}$ of the anti-Stokes photon 
using a standard Hanbury-Brown-Twiss (HBT) interferometer \cite{3}. 
The experimental set up of this measurement is shown in Fig.~\ref{supfig2}. 
The beam splitter (BS) used here has a half-half splitting ration with 
T and R denoting the transmission and reflection ports. 
D1, D2 and D3 are three single photon detectors that record 
the click event triggered by the photon in channel G, T and R, respectively.
Definition of the conditional second-order auto-correlation function is \cite{3}
\begin{eqnarray}
\begin{array}{ll}
&g_c^{(2)}=P(TR|G)/[P(T|G)P(R|G)]
\end{array}
\end{eqnarray}
where $P(R|G)$, $P(T|G)$ and $P(TR|G)$ are the conditional events probability, after the detector D1 click, of detector D2 click, detector D3 click and both detectors click at the same time, respectively. With the relation that $P(X|G)=N(X)/N(G)$, we have the following relation:
\begin{eqnarray}
\begin{array}{ll}
&g_c^{(2)}=N(G)N(GTR)/[N(GT)N(GR)]
\end{array}
\end{eqnarray}
where $N(G)$,$N(GR)$,$N(GT)$ and $N(GTR)$ are D1 click counts, D1-D2 2-photon coincidence counts, D1-D3 2-photon coincidence counts and D1-D2-D3 3-photon coincidence counts, respectively. Here we can see that for anti-Stokes photon in ideal single photon state we have $g_c^{(2)}=0$ and in the pure 2-photon state we have $g_c^{(2)}=0.5$, thus in reality we may require the final $g_c^{(2)}<0.5$ as the criterion for the single photon source.

For a anti-Stokes photon, after sent onto a 50/50 beam splitter, it can be transmitted or reflected into the path T and path R with equal length and then finally be detected by single photon detectors placed in the transmission (D2) and reflection path (D3). Here, when there is a click in the Stokes photon channel(D1), the photon click events in both D2 and D3 with the time lapse ($t_T$ and $t_R$) relative to the click of Stokes photon will be recorded. To measure the $g_c^{(2)}$ at zero delay, $t_T=t_R$ must be hold in our time resolution. If at $t_T=t_R$, a coincidence count was recorded between D2 and D3, we treat it as a non-single photon event. 

In reality, however, to make this measurement more safe, any coincidence count within a reasonable window defines as $\Delta{t}_w > |t_T-t_R|$ will be treated as non-single photon. Therefore, in the Fig.(2) of main text, the $g^{(2)}_c$ with different $t_w$ are measured and presented to show the good single photon quality of both the photons from the source and the photons retrieved from the storage.
\\

\subsection{Waveform likeness} 

\begin{figure}[htbp]
\centering
\fbox{\includegraphics[width=8.5cm]{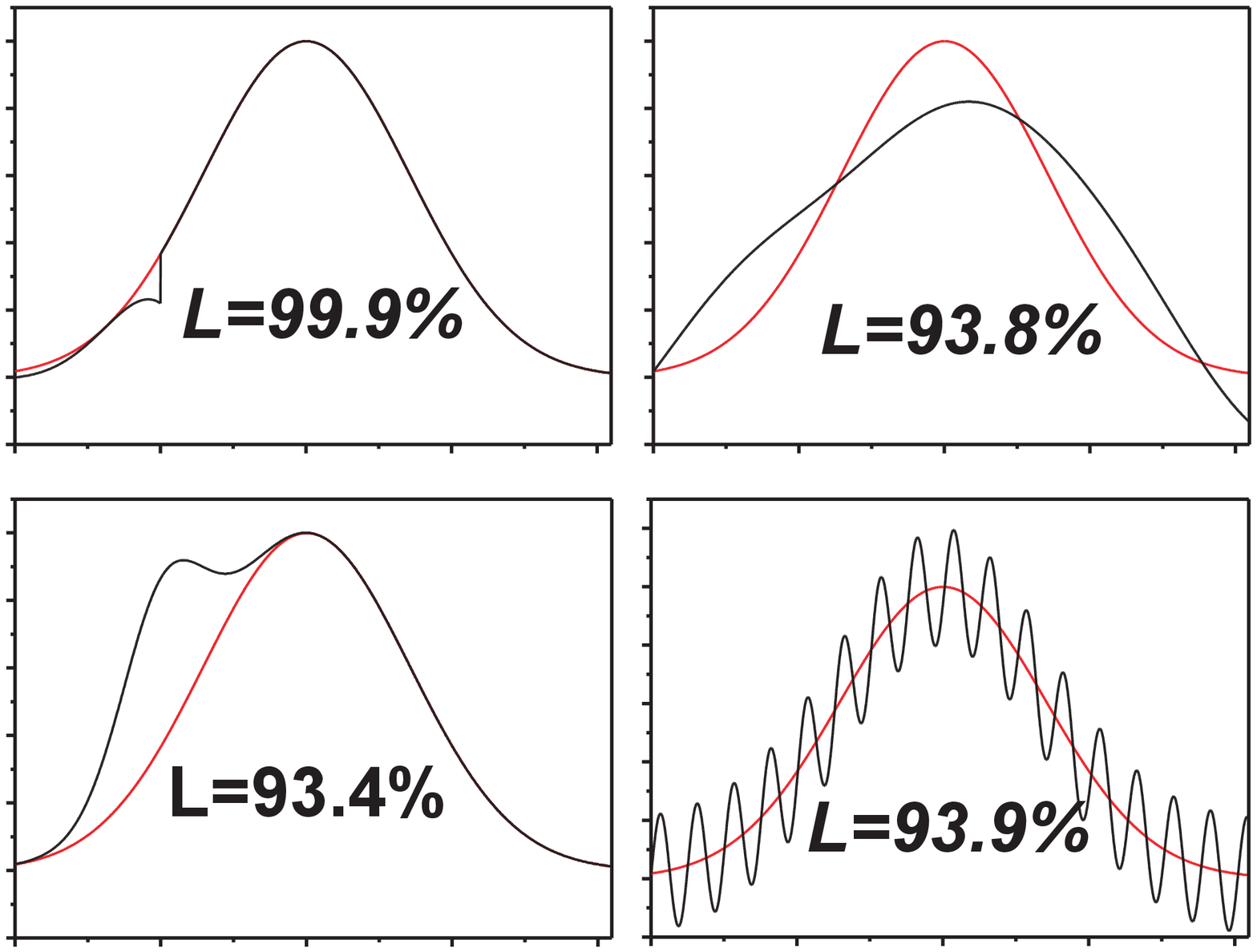}}
\caption{ Fake curves shown how senstively does the defined likeness depend on the waveform shape .}
\label{supfig3}
\end{figure}

\noindent
The classical fidelity \cite{EIT2013Yu} is defined by
\begin{eqnarray}
\begin{array}{ll}
&F_{class}=\frac{|{\int}E_{in}^\ast(t-t_d)E_{out}(t)dt|^2}{[\int|E_{in}(t)|^2dt][\int|E_{out}(t)|^2dt]}
\end{array}
\end{eqnarray}
Similarly, for a quantitative estimation, we calculate the temporal waveform likeness with the following definition:
\begin{eqnarray}
\begin{array}{ll}
&L=\frac{|{\int}\psi_{in}^\ast(\tau-\tau_{delay})\psi_{out}(\tau)d\tau|^2}{[\int|\psi_{in}(\tau)|^2d\tau][\int|\psi_{out}(\tau)|^2d\tau]}
\end{array}
\end{eqnarray}

According to Ref\cite{4,5}, 
$\psi_{in}(\tau)=\sqrt{G_{in}^{(2)}(\tau)}$ and 
$\psi_{out}(\tau)=\sqrt{G_{out}^{(2)}(\tau)}$ are 
reasonable approximations with $G_{in}(\tau)$ and $G_{out}(\tau)$ 
are Glauber correlation function before and after storage.
Therefore, with waveform likeness $L$ can be defined as :
\begin{eqnarray}
\begin{array}{ll}
&L=\frac{|{\sum}\sqrt{N_{in}(\tau-\tau_{delay}) N_{out}(\tau)}|^2}
{\sum|\sqrt{N_{in}(\tau-\tau_{delay})}|^2 \times \sum|\sqrt{N_{out}(\tau-\tau_{delay})}|^2}
\end{array}
\end{eqnarray}

with $N_{in/out}(\tau)=\eta_{channel} G_{in/out}^{(2)}(\tau) \Delta t_{bin} T$.
Here $\eta_{channel}$ is the total channel efficiency,
$\Delta{t_{bin}}$ is the time bin and 
$T$ is the total photon generation time. 
To give an intuitive impression about this value, we show 
a few fake curves in Fig.\ref{supFig3} that testing how sensitive this quantity can 
be used to tell the waveform similarity.

\end{document}